%% file: sample-manuscript.tex
\documentclass[acmsmall]{acmart}
\usepackage{multirow}
\AtBeginDocument{%
  }

\setcopyright{acmlicensed}
\acmJournal{PACMHCI}
\acmYear{2025} \acmVolume{9} \acmNumber{7} \acmArticle{439} \acmMonth{11}\acmDOI{10.1145/3757620}

\begin{document}

\title[Learning AI Auditing]{Learning AI Auditing: A Case Study of Teenagers Auditing a Generative AI Model}


\author{Luis Morales-Navarro}
\email{luismn@upenn.edu}
\orcid{0000-0002-8777-2374}
\affiliation{%
  \institution{University of Pennsylvania}
  \city{Philadelphia}
  \state{Pennsylvania}
  \country{USA}
}

\author{Michelle Gan}
\email{michellegan00@gmail.com}
\orcid{0009-0000-6632-0021}
\affiliation{%
  \institution{University of Pennsylvania}
  \city{Philadelphia}
  \state{Pennsylvania}
  \country{USA}
}

\author{Evelyn Yu}
\email{evelynyu@upenn.edu}
\orcid{0009-0000-1655-6116}
\affiliation{%
  \institution{University of Pennsylvania}
  \city{Philadelphia}
  \state{Pennsylvania}
  \country{USA}
}

\author{Lauren Vogelstein}
\email{lev2124@tc.columbia.edu}
\orcid{0000-0002-2317-2513}
\affiliation{%
  \institution{Teachers College, Columbia University}
  \city{New York}
  \state{New York}
  \country{USA}
}

\author{Yasmin B. Kafai}
\email{kafai@upenn.edu}
\orcid{0000-0003-4018-0491}
\affiliation{%
  \institution{University of Pennsylvania}
  \city{Philadelphia}
  \state{Pennsylvania}
  \country{USA}
}

\author{Danaé Metaxa}
\email{metaxa@upenn.edu}
\orcid{0000-0001-9359-6090}
\affiliation{%
  \institution{University of Pennsylvania}
  \city{Philadelphia}
  \state{Pennsylvania}
  \country{USA}
}

\renewcommand{\shortauthors}{Luis Morales-Navarro, et al.}

\begin{abstract}

This study investigates how high school-aged youth engage in algorithm auditing to identify and understand biases in artificial intelligence and machine learning (AI/ML) tools they encounter daily. With AI/ML technologies being increasingly integrated into young people's lives, there is an urgent need to equip teenagers with AI literacies that build both technical knowledge and awareness of social impacts. Algorithm audits (also called AI audits) have traditionally been employed by experts to assess potential harmful biases, but recent research suggests that non-expert users can also participate productively in auditing. We conducted a two-week participatory design workshop with 14 teenagers (ages 14–15), where they audited the generative AI model behind TikTok's Effect House, a tool for creating interactive TikTok filters. We present a case study describing how teenagers approached the audit, from deciding what to audit to analyzing data using diverse strategies and communicating their results. Our findings show that participants were engaged and creative throughout the activities, independently raising and exploring new considerations, such as age-related biases, that are uncommon in professional audits. We drew on our expertise in algorithm auditing to triangulate their findings as a way to examine if the workshop supported participants to reach coherent conclusions in their audit. Although the resulting number of changes in race, gender, and age representation uncovered by the teens were slightly different from ours, we reached similar conclusions. This study highlights the potential for auditing to inspire learning activities to foster AI literacies, empower teenagers to critically examine AI systems, and contribute fresh perspectives to the study of algorithmic harms.
\end{abstract}




\received{October 2024}
\received[revised]{April 2025}
\received[accepted]{August 2025}
\maketitle

\input{sections/1-intro}
\input{sections/2-background}

\input{sections/3-methods}
\input{sections/4-findings}

\input{sections/5-discussion}
\input{sections/6-conclusion}

\begin{acks}
With regards to Lucia Kulzer, Alexis Cabrera-Sutch, Carly Netting, and Danielle Marino for their support during the participatory design workshop. This study was supported by National Science Foundation (NSF) grants \#2333469 and \#2342438. Any opinions, findings, and conclusions or recommendations expressed in this paper are those of the authors and do not necessarily reflect the views of NSF, the University of Pennsylvania, or Columbia University.
\end{acks}

\bibliographystyle{ACM-Reference-Format}
\bibliography{sample-base}
\pagebreak 
\begin{appendix}

\section{Coding Scheme}

\begin{table}[h]
\caption{Coding scheme used in our analysis of the dataset generated by youth.}
\resizebox{\columnwidth}{!}{
\begin{tabular}{lll}
\toprule
\textbf{Axis}                    & \textbf{Category}                                & \textbf{Codes}                                                                                                                                                                                                                                         \\ \hline
\multirow{7}{*}{\textit{Gender}} & \multirow{3}{*}{Change in gender representation} & \begin{tabular}[c]{@{}l@{}}Yes: gender represented in the input image is different than \\ gender represented in the output image (e.g., input has a \\ masculine-presenting figure and output image has a \\ feminine-presenting figure)\end{tabular} \\ \cline{3-3} 
                                 &                                                  & \begin{tabular}[c]{@{}l@{}}No: gender represented in the output image is the same as in \\ the input image\end{tabular}                                                                                                                                \\ \cline{3-3} 
                                 &                                                  & \begin{tabular}[c]{@{}l@{}}Ambiguous: it is not clear if there was a change in gender \\ representation; it is not clear if output image is \\ masculine-presenting or feminine-presenting\end{tabular}                                                \\ \cline{2-3} 
                                 & \multirow{2}{*}{Facial Hair*}                    & Presence: visible facial hair, including stubble                                                                                                                                                                                                       \\ \cline{3-3} 
                                 &                                                  & Absence: no visible facial hair                                                                                                                                                                                                                        \\ \cline{2-3} 
                                 & \multirow{2}{*}{Blush or Mascara*}               & Presence: visible blush or mascara                                                                                                                                                                                                                     \\ \cline{3-3} 
                                 &                                                  & Absence: no visible blush or mascara                                                                                                                                                                                                                   \\ \hline
\multirow{4}{*}{\textit{Age}}             & \multirow{2}{*}{Wrinkles}                        & Presence: visible wrinkles, including smile lines                                                                                                                                                                                                      \\ \cline{3-3} 
                                 &                                                  & Absence: no visible wrinkles                                                                                                                                                                                                                           \\ \cline{2-3} 
                                 & \multirow{2}{*}{Gray Hair}                       & Presence: visible gray hair                                                                                                                                                                                                                            \\ \cline{3-3} 
                                 &                                                  & Absence: no visible gray hair                                                                                                                                                                                                                          \\ \hline
\multirow{8}{*}{\textit{Race}}            & \multirow{3}{*}{Change in racial representation} & \begin{tabular}[c]{@{}l@{}}Yes: race represented in the input image is different than race \\ represented in the output image (e.g., input has an Asian man \\ and output as a White man)\end{tabular}                                                 \\ \cline{3-3} 
                                 &                                                  & \begin{tabular}[c]{@{}l@{}}No: race represented in the output image is the same race as in \\ the input image\end{tabular}                                                                                                                             \\ \cline{3-3} 
                                 &                                                  & \begin{tabular}[c]{@{}l@{}}Ambiguous: it is not clear if there was a change in racial \\ representation\end{tabular}                                                                                                                                   \\ \cline{2-3} 
                                 & \multirow{2}{*}{Fade**}                          & \begin{tabular}[c]{@{}l@{}}Presence: visible fade hairstyle hair on the sides of head is \\ shorter than the hair on the top of the head\end{tabular}                                                                                                  \\ \cline{3-3} 
                                 &                                                  & Absence: no visible fade hairstyle                                                                                                                                                                                                                     \\ \cline{2-3} 
                                 & \multirow{3}{*}{Hairstyle changes**}             & Curlier: output image has curlier hair than input image                                                                                                                                                                                                \\ \cline{3-3} 
                                 &                                                  & Straighter: output image has straighter hair than input image                                                                                                                                                                                          \\ \cline{3-3} 
                                 &                                                  & \begin{tabular}[c]{@{}l@{}}Same: there is no difference in hairstyles between input and\\ output images.\end{tabular}                                                                                                                                  \\ \hline
\multicolumn{3}{l}{\begin{tabular}[c]{@{}l@{}}* included in our analysis to examine youth’s use of these features as proxies for gender\\ ** included in our analysis because some youths noted that these could be racially coded features\end{tabular}}                                                                                   
\end{tabular}}
\end{table}
\end{appendix}

\end{document}

%% file: sections/1-intro.tex
\section{Introduction}
% Motivation
Artificial intelligence and machine learning (AI/ML) technologies have become deeply woven into young people's daily lives over the past ten years---from photo filters on social media to recommendations on streaming platforms to voice assistants. This widespread adoption of AI/ML creates an urgent need to support teenagers in developing AI literacies~\cite{long2020ai}. Young people must gain knowledge not only about how these technologies function but also about their potential societal impacts. Most critically, teenagers need the skills to identify algorithmic biases and take action when these systems cause harm. While research has shown that teenagers are able to identify algorithmic bias and harms~\cite{solyst2023potential, morales2024youth}, translating these insights into approaches to further systematic and empirical examinations of algorithmic systems remains an open challenge. As~\citet{solyst2023potential} note, we lack sufficient studies examining how to cultivate and harness young people's valuable insights for identifying and reducing algorithmic harm. This gap reflects a broader tendency to underestimate teenagers' capacities to grasp both the technical complexities and the ethical implications of AI/ML technologies, leading instead to attempts to limit their engagement ~\cite{solyst2024children}. 

% Related work
One efficient method that experts have developed to examine and draw conclusions about AI/ML systems, particularly in regard to potential harmful biases and discrimination, is algorithm auditing~\cite{metaxa2021auditing}. More recently, scholars have begun to examine how everyday people can engage in algorithm auditing, reframing auditing as a way for the public to gain insights about algorithmic behaviors in everyday contexts~\cite{shen2021everyday}. In this paper, we explore the potential for engaging teenagers in full-fledged auditing activities to investigate potentially harmful algorithmic biases on a popular social media platform.

% Method
We present a descriptive case study of a participatory design workshop in which teenagers engaged in auditing the generative AI model that powers TikTok filters. This model is accessible through Effect House, a filter development environment that supports the creation of text-to-image filters, the same filters teens encounter on TikTok.  We conducted the workshop with a group of 14 teens (ages 14–15) in a two-week summer program. Workshop activities were designed to support participants in systematically investigating potentially harmful algorithmic biases. Rather than becoming experts in auditing, our intention in this workshop was for young people to lead the process of evaluating an algorithmic system they encounter on a daily basis. In addition to describing participants' experiences throughout the auditing process, we also triangulated their findings to examine if the workshop design supported participants to identify potentially harmful algorithmic biases in their audit. 
%contextualize and validate their process by comparing their observations, approaches, and findings with those that would result from a more formal, expert-led audit. 
We address the following research questions: (1) How did participants engage in this algorithm auditing activity? In particular, what choices did they make, how did these choices vary across participants, and what reflections did they have throughout the process? and (2) Did the workshop support participants to reach evidence-based, credible conclusions in their audit? 
 %And to contextualize and validate their audit process and findings, (2) How would each step in the youths' audit compare with more traditional, expert-led audit procedures? 

% Findings
Our case study demonstrates that with adequate scaffolding, teenagers can participate in full-fledged audits of real-world algorithmic systems they encounter in their daily lives. We observed participants making connections to their everyday experiences and understandings of biases; for example, they selected inputs based on the social dynamics they observed in their own communities and contributed ideas about age-related biases that are uncommon in professional audits but were particularly salient to them. By triangulating participants' findings, we were able to confirm that the workshop design may be conducive to supporting teens to conduct audits with evidence-based, credible conclusions.

%Even though the youths' exact numerical findings differed from our own analysis, we were able to validate their overall conclusions by conducting our own expert analysis, confirming that this activity led them to realistic and accurate conclusions.

% Although researchers and youths obtained different results, the conclusions drawn from both analyses were similar, namely that—in terms of gender representation and biases—Effect House's filter masculinized input images  when prompted with occupations like rapper, carpenter, priest, and janitor, and feminized outputs for occupations like fast food workers, receptionists, and nail technicians. 

% Discussion
We discuss the design of audit-based learning activities and the role that these can play in supporting the development of AI literacies for young people. Further, we explore this case study as confirming the value of involving non-experts in auditing algorithmic systems with which they are personally familiar. Our paper makes the following contributions: (1) we illustrate one process by which teenagers engage in a scaffolded auditing learning activity to empirically investigate potentially harmful behaviors in a real-world algorithmic system, (2) we provide evidence from a case study about the deployment of one such activity in practice, and (3) we reflect on participants' experiences with the activity, situating their approaches, decisions, and conclusions in relation to more traditional and participatory auditing procedures.

%% file: sections/2-background.tex
\section{Background \& Related Work}
\label{sec:relwork}
In this section, we position this study in relation to other studies about teenagers and algorithmic justice, algorithm auditing as a whole, and participatory approaches to auditing. We also provide relevant background on the specific domain our teen participants audited: identity representation of different occupations as reflected through the lens of algorithmic systems---a popular domain for prior expert audits.

\subsection{Teens and Algorithmic Justice}
In the last ten years, computing education and CSCW researchers have recognized the importance of engaging teenagers in learning activities that investigate ``the consequences, limitations, and unjust impacts of computing in society''~\cite{ko2020time}. These activities typically involve reframing computing as a sociotechnical field by prompting learners to explore the functionality of computing systems and their implications~\cite{morales2023conceptualizing, solyst2023potential}. Such activities are often designed for young people to assess inputs and outputs of computing systems, investigate how systems are actually used, and consider how they affect people and the environment~\cite{dindler2023dorit}. However, most efforts to engage teenagers in being critical about computing tend to focus on discussion or direct instruction without providing opportunities for learners to empirically investigate issues of justice and ethics in computing~\cite{solyst2023potential, morales2024unpacking}. 

Limited research has examined how young people actively investigate issues of algorithmic justice---how they understand, explain, and investigate the potential for algorithmic systems to perpetuate harm~\cite{birhane2021algorithmic}. Researchers have used participatory design (PD) workshops to study how young people think about these issues~\cite{coenraad2022s, solyst2024children}. For instance, studies have found that while teenagers may be aware of the negative impacts of technology in their everyday lives, they may not use the word ``bias''  ~\cite{coenraad2022s}. Other work has highlighted that teenagers may view AI/ML systems as being bias-free or view bias favorably when it enhances their own user experience and negatively when it restricts it ~\cite{lee2022eliciting}. \citet{salac2023funds} noted that when evaluating scenarios in which issues of algorithmic justice were presented, teenagers considered their own lived experiences, how systems may behave in different contexts, and larger societal issues that mediate the use of the systems. While these studies investigate young people's perceptions of issues related to algorithmic justice, they do not address how young people can be engaged in investigating these issues themselves.  

\subsection{Algorithm Auditing}
This paper addresses a gap in the literature on teenagers and algorithmic justice by exploring how algorithm auditing activities may be used to engage teenagers in empirical investigations of algorithmic justice issues. Auditing algorithmic systems involves ``repeatedly querying an algorithm and observing its output in order to draw conclusions about the algorithm's opaque inner workings and possible external impact''~\cite{metaxa2021auditing}. In contrast to other forms of evaluation, auditing aims to draw system-wide conclusions rather than scoping its conclusions to a specific set of test cases. Additionally, audits are often external evaluations conducted by third parties without insider access or knowledge. Audits are traditionally conducted by experts seeking to evaluate how AI/ML-powered systems behave across different application areas (housing, healthcare, social media, search) and draw inferences about the potential harmful impact of these systems (for a review of expert algorithm audits, see~\cite{bandy2021problematic}). Although the specific procedures are tailored to each audit, auditing usually involves (1) developing a hypothesis about a specific system behavior, (2) generating a set of systematic, thorough, and thoughtful inputs to test the hypothesis, (3) running the tests, (4) analyzing the data, and (5) reporting the results (for more details on the method, see~\cite{metaxa2021auditing}).

\subsection{Participatory Approaches to Auditing}
Engaging relevant communities in cooperative and participatory design (PD) activities to design and evaluate computing systems has a long tradition in CSCW research ~\cite{zacklad2003communities, bannon2012design}. Recently such approaches have been adopted in efforts to design and evaluate AI/ML systems in community-engaged ways \cite{queerinai2023queer, suresh2022towards}.
Such PD activities can be structurally open or closed with participants having different levels of autonomy to define their engagement ~\cite{zacklad2003communities}.

Researchers have begun exploring the potential for non-experts to engage in identifying potentially harmful algorithmic behaviors through auditing-like practices. This can be done under frameworks known as \textit{crowdsourced}, \textit{everyday}, or \textit{end-user audits}~\cite{shen2021everyday,lam2022end,sandvig2014auditing}. In some cases, researchers have provided non-expert adults with user-friendly interfaces to scaffold auditing tasks that harness users' personal experiences~\cite{lam2022end}, while in others users contribute data without actively participating in the larger auditing endeavor~\cite{lam2023sociotechnical}. Other literature observes that users may sometimes engage organically with auditing practices by evaluating algorithmic systems in their everyday lives in the absence of experts~\cite{shen2021everyday, devos2022toward}.

\subsubsection{Teens and Algorithm Auditing}

Several efforts have started to explore different participatory approaches to engage teenagers in auditing-related activities. \citet{solyst2023potential} adapted user-driven everyday auditing tasks, such as making observations of Google search results, into activities for middle schoolers. In their study, researchers presented participants with bias identification scenarios from Google search and DALL-E-generated images, finding that most participants identified race-related biases as harmful. Participants in this study also argued that users should have the ability to report problematic behaviors and that AI systems should warn users of potential harm. \citet{morales2024youth} designed a workshop in which high school participants designed ML-powered physical computing projects and then audited their peers' projects. They found that when externally evaluating motion classifiers, participants were able to identify unexpected biases. In doing so, participants also inferred dataset and model design issues that could cause such biases. However, the auditing activities in these studies primarily focused on having participants make a single or a few observations about a system's behavior, rather than the systematic analysis of inputs and outputs and the full process of conducting an audit from beginning to end. We expand on this research by engaging teenagers in a full-fledged audit of a real-world algorithmic system, specifically investigating how teenagers engaged in auditing gender and racial representation in the outputs of the generative image model of TikTok's Effect House. \citet{iversen2017child} argue that young people can also engage in such participatory activities by becoming \textit{protagonists} or the main decision-makers. In our study, we positioned participants as protagonists, enabling them to decide what to audit and how to analyze their data.

It is worth noting that this kind of PD work does not happen in a vacuum, as researchers set parameters and create basic structures that orient participant engagement ~\cite{zacklad2003communities}. The role of researchers in such activities is that of encouraging and supporting young people to “be the main agents in driving the design process and thereby to develop skills to design and reflect on technology and its role in their lives” \cite{iversen2017child}. Here, researchers can be \textit{reflective practitioners} that analyze the design process and its outputs \cite{bannon2012design, schon2017reflective}. When PD is related to a learning or educational intervention, reflecting on the design process often involves examining if the PD sessions are conducive to learning ~\cite{paudel2024leveraging, bossen2016evaluation}. This is particularly important as PD could lead to misinformation and disinformation, presenting  risks "by platforming false or hateful ideas and allowing them to gain impact” ~\cite{schafer2023participatory}. As such, in our study we triangulate participants' findings in an effort to examine if the design of the PD workshop supported teens to reach evidence-based, credible conclusions. 

\subsection{Representation of Occupations in Algorithmic Systems}
In our study, teenage participants conducted an audit of identity representation in occupations. As we will describe in a later section, they came to this goal after their own independent explorations with the tool. Notably, this is a popular domain for prior audits; several expert and non-expert audits have investigated gender and racial representation of occupations in image search and image generation.

\subsubsection{Expert Audits on Representation in Occupations}

About a decade ago, \citet{kay2015unequal} conducted an audit on gender representation in Google image search results for common occupations (e.g., ``doctor'', ``engineer'') that found systematic underrepresentation of women. A follow-up study by  \citet{metaxa2021auditing} investigated gender and racial representation for image search results, again finding evidence of men's systematic overrepresentation, as well as the overrepresentation of White people in image search results. These two studies established common methodologies for  conducting algorithm audits of gender and racial representation of occupations that have been followed and replicated in studies of generative AI image models. Such studies have shown systemic amplification of racial and gender disparities in the representation of occupations in image outputs across different models (Dall-E2, Stable Diffusion v1.4 and v2) ~\cite{naik2023social,bianchi2023easily,luccioni2024stable}.

\subsubsection{Adult and Teen End-user Audits on Representation in Occupations}

Building on expert audit research, \citet{devos2022toward} investigated how non-expert adults identified harmful behaviors in algorithmic systems such as Google image search. They conducted interviews in which users were tasked to search for words such as ``librarian'' or ``thug'' to prompt them to explain how they thought about potential harmful biases in search results. Following, when asked to look for other cases of harmful algorithmic behaviors, participants conducted their own searches, observing racial and gender biases in the representation of occupations such as ``computer scientist'', ``maid'', and ``firefighter.'' This study found that users' experiences and exposure to societal biases influence their strategies for conducting image searches and how they interpret the results.

% \subsubsection{Youth Audits and Representation of Occupations}

These tasks have been adapted to study how teenagers engage in auditing-like practices. For example, \citet{solyst2023potential} conducted a PD workshop where teenagers analyzed search results for images of computer programmers and Dall-E-generated images of doctors. \citet{morales2024youth} used similar tasks in a pre/post interview study to assess how teenagers' identification of potential
algorithmic biases and harms changed after participating in peer-auditing activities.
These studies found that teenagers were able to recognize and explain potentially harmful algorithmic biases in the representation of occupations and that teenagers were concerned that the stereotypes in the representation of occupations may discourage young people from pursuing certain careers. Furthermore, these two studies highlight the potential of involving teenagers as contributors in the evaluation of algorithmic systems that they use in their everyday lives. 

While previous studies engage teens in auditing-like tasks, they stop short of supporting teens in conducting end-to-end, full-fledged algorithm audits. Our study addresses this gap by presenting a case study in which teenagers investigate gender, race and age biases in the representation of occupations in  the generative image model used in TikTok filters.

%% file: sections/3-methods.tex
\section{Methods}

In this section, we describe the participants and context of our study, our data collection and analyses processes, and our research team's positionality in conducting this work.

\subsection{Participants}

This study was conducted in the context of a series of iterative participatory design workshops with teenagers that investigated the potential of developing algorithm auditing learning activities (for details on other iterations of this work, see \cite{vogelstein2025rapid, morales2025learning}). We worked with teenagers enrolled in a four-year afterschool program, STEM Stars (a pseudonym), at a science center in the Northeastern United States. As part of the STEM Stars program, we provided workshops on algorithm auditing in the fall, spring, and summer of 2024. Participants were invited to take part in the study through emails and texts sent out by the science center staff. Guardians filled out consent forms before their children participated in the study, and minors assented to participate. The Institutional Review Board of the University of Pennsylvania approved the study protocol. The analysis in this paper focuses on our two-week summer workshop with 14 teens (14-15 years old) in the STEM Stars program. That summer, we worked with six female, one non-binary, and seven male youth. The majority were from marginalized racial backgrounds, with all but one identifying as African American, Asian American, or multiracial. To respect the privacy of our underage participants, all names used in this paper are pseudonyms.

\subsection{Context}

In this paper, we focus on how participants audited the generative AI model that powers TikTok's filters, which are created in an application called Effect House. We provide context on TikTok and Effect House below.

\subsubsection{TikTok Filters}
TikTok is a prominent video-sharing social media platform where users create and share short videos. TikTok enables users to record, edit, and remix short videos, which can include generative AI-driven effects or filters. The platform is particularly popular among teenagers, with 58\% of teens in the United States reporting using TikTok on a daily basis~\cite{anderson2023teens}. Recent documents from court cases show that TikTok estimates that 95\% of teens under 17 who have a smartphone use TikTok~\cite{allyn24}. Previous research indicates that the app's popularity may be influenced by the strength of its recommendation system~\cite{zhao2020analysis}. This is particularly relevant for youth, as 17\% of teens in the US describe themselves as being on TikTok almost constantly. At the same time, regulators argue that the success of the recommendation system can have a noxious effect on teenagers through compulsive usage~\cite{allyn24}. While the details of the recommendation system are not public, users build hypotheses about the system by testing it themselves or replicating collective memes shared on social media~\cite{karizat2021algorithmic}.
 %Despite its popularity, TikTok has not been widely audited. In one audit, researchers used sockspuppet accounts to show that recommendations are influenced by user characteristics including language, geography, and behavior~\cite{boeker2022empirical}. 

\begin{figure}[ht]
\centering
\includegraphics[width=0.6\textwidth, ]{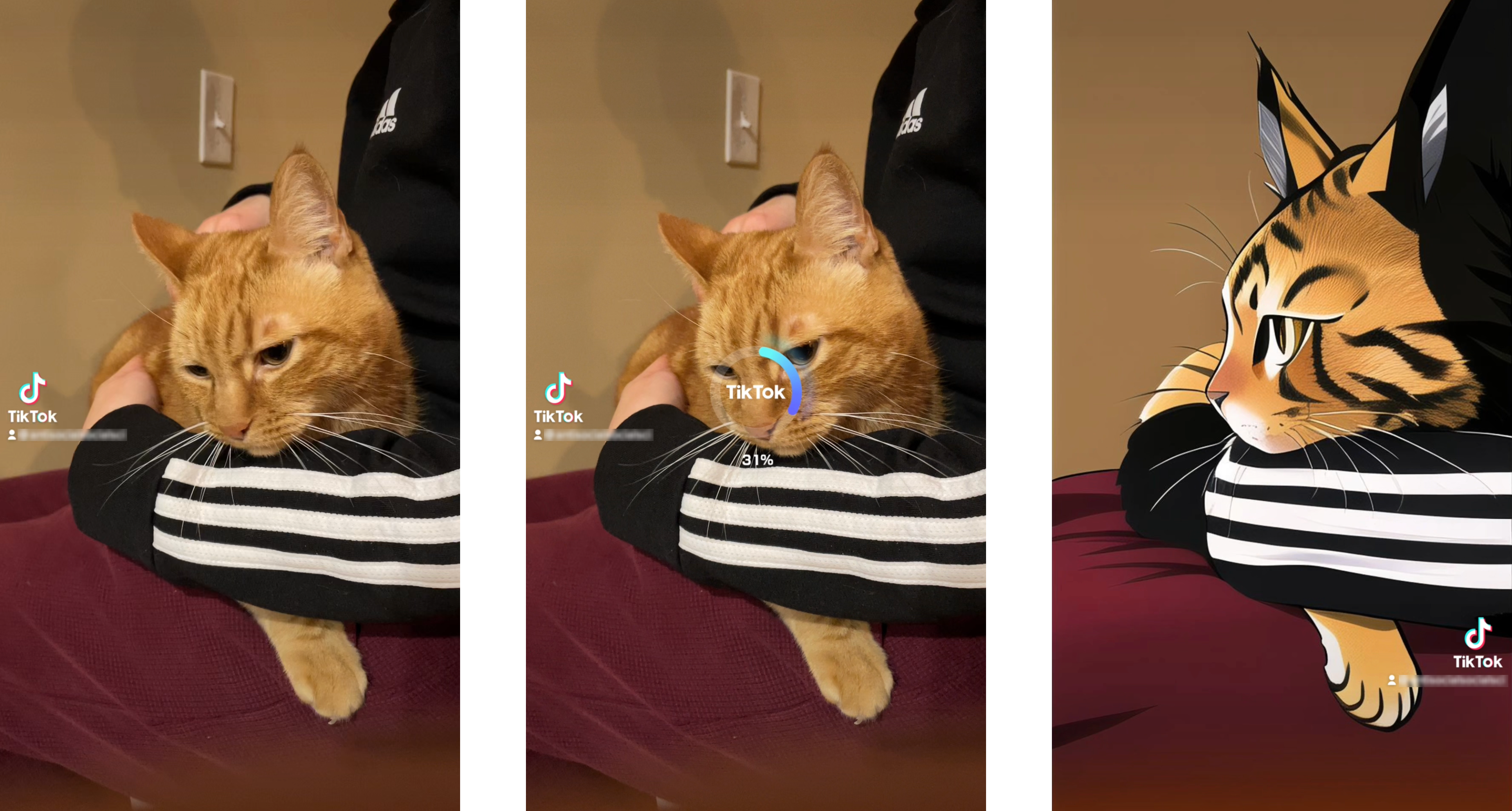}
\Description[Three screenshots showing a cat photo being turned into a manga-style illustration.]{3 screenshots. Left: photo of an orange cat. Center: same photo with a progress circle loading. Right: manga-style orange cat with tiger stripes.}
\caption{Example of a TikTok filter in action. This filter uses generative AI to output a manga-style illustration of the input photo.}
\label{fig:carl}
\end{figure}

Generative AI filters on TikTok are text-to-image or image-to-image filters that use generative AI models to modify users' photo and video inputs. Both the original and altered content can then be included in the posted video. For an illustration of a filter in action, see Figure \ref{fig:carl}. The mechanisms behind these filters---and the underlying models that power them---are largely opaque to both users and experts. This lack of transparency raises concerns, as evaluation work of generative AI models at large (outside TikTok) has demonstrated that these models can replicate stereotyped gender attributes~\cite{mannering2023analysing, sun2023smiling, zhang2023auditing} and societal biases~\cite{naik2023social}. Other problematic behaviors include the oversexualization of features and the promotion of unrealistic beauty standards~\cite{bonner2023filters}. 

Research on such filters on other platforms has shown that biased AI-generated changes of facial features can lead to negative self-perception~\cite{burnell2022snapchat, pescott2020wish} and may increase body dysmorphia among teens~\cite{laughter2023psychology}. However, the design of such filters is largely unregulated~\cite{doh2024pixels}. Although TikTok does publish guidelines for filter designers that explicitly prohibit stereotypes and discrimination~\cite{guidelines24}, accountability mechanisms available to users are few and opaque~\cite{doh2024pixels}.

\subsubsection{Effect House Interface}
\begin{figure}[ht]
  \centering
  \includegraphics[width=\linewidth]{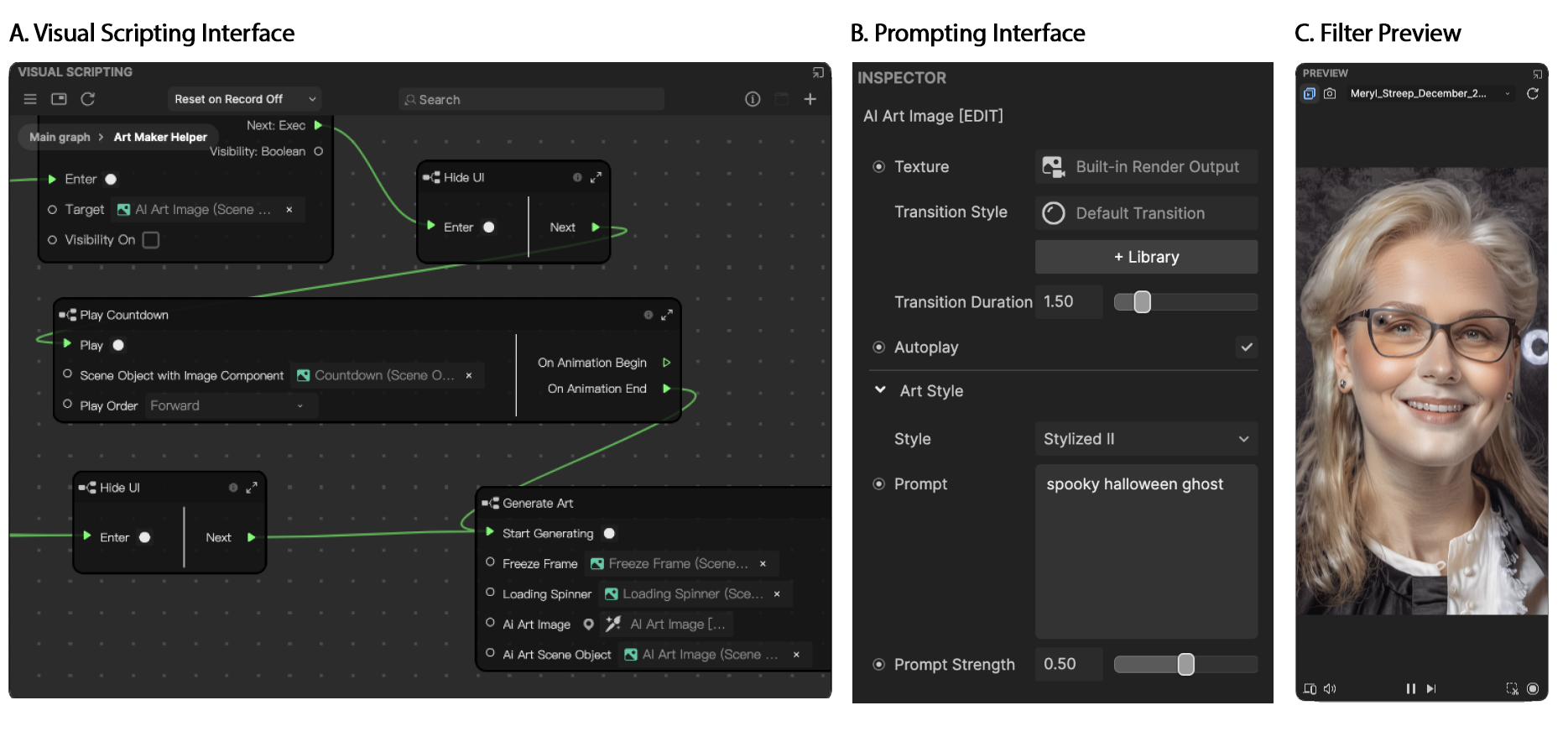}
  \caption{Effect House's visual scripting, prompting, and filter preview interfaces.}
  \Description{Three screenshots of the Effect House interface, including a block-based visual scripting interface, an action menu-based prompting interface, and a filter preview of the output.}
  \label{fig:effect}
\end{figure}

The Effect House development environment includes a visual scripting interface for filter creators to write code (see Figure \ref{fig:effect}.a), and also provides a prompting interface (Figure \ref{fig:effect}.b) that enables designers to create filters by writing their own text prompts, which are used to run the default generative AI model provided by TikTok. 

% on generative AI-powered TikTok filters, exploring how teens audited the default generative model used in Effect House, TikTok's filter development environment. 
% Generative AI filters on TikTok are text-to-image or image-to-image filters that use generative AI models to modify users' photo inputs. 
% Both the original and altered content can then be included in the posted video. 
% For an illustration of a filter in action, see Figure X. 
% Effect house includes a visual scripting interface (see \dm{Figure X.a}) and an AI Texture interface that enables designers to create stylized filters through text prompts (see Figure X.a).
Relevant to this study are three key parts of the Effect House interface: (1) the input text prompt, where filter designers can write prompts for their filters, and (2) the filter preview, where designers can input images (Effect House also has some built-in options) to preview an output image that Effect House's generative AI model creates based on the prompt and input image.
The interface also includes some togglable parameters, like the ``prompt strength'' slider (a 0-1 value set to 0.5 by default) that, when increased, exaggerates the degree to which the input image is stylized.

% HERE EXPLAIN THE PROMPT STRENGTH THING
 
\subsection{Workshop Procedure}

%\subsubsection{Workshop Procedure}
During the workshop, we met with participants for 28 hours over the course of two weeks. All participants had previously participated in a short 4-hour workshop in which they informally audited an anime TikTok filter (a real filter available to all users on TikTok). In the first week, teens designed generative AI TikTok filters and audited each other's filters. Even though they were able to systematically evaluate the filters, the number of tests conducted by most pairs of participants was limited (with each group running an average of 30 tests), which impacted their abilities to draw conclusions based on strong evidence. In the second week, we shifted focus to auditing the Effect House tool itself. In this paper, we focus on that second week's activities in which participants collaboratively audited the default generative model built into Effect House. To address the problem of data scale observed in week one, we decided to have participants conduct the audit collaboratively, developing a hypothesis, generating a set of prompt inputs for the audit, and testing prompts as a group. This enabled participants (with the support of researcher-facilitators) to generate 100 prompts that were used in conjunction with 12 input images provided by researchers to run 1200 tests, creating a more robust dataset for analysis. After creating the dataset, participants analyzed the data in small groups of 2-3, with each group producing an audit report to present to the wider group.

Workshop activities were scaffolded around five steps: (1) developing a hypothesis, (2) generating inputs, (3) running the tests, (4) analyzing the results, and (5) creating an audit report (see Table \ref{tab:activities}). Every workshop day started with a 30-minute game or warm-up activity (e.g., duck duck goose) and included two snack breaks. Since participants had already participated in auditing activities in the first week, there was little to no formal instruction during the second week. Instead, researcher-facilitators provided opportunities for participants to review how audits are conducted and brainstorm ideas for each activity step. Each day was scheduled as follows (see Table \ref{tab:activities} for more details):

\paragraph{Day 1} The main researcher-facilitator led an activity for participants to review the auditing process. Afterwards, teens spent time brainstorming hypotheses about the Effect House generative AI model.  To get them started in the process of hypothesis formation, we asked participants to spend an hour in small groups trying different prompts and inputs in Effect House and to take notes of observed behaviors they might want to investigate. They shared their hypotheses, and then the main researcher-facilitator proposed a hypothesis that encompassed several ideas presented by the youth: Effect House's image model reinforces gender and race stereotypes about different occupations.

\paragraph{Day 2} Participants began the day brainstorming different ways to test their hypothesis.  The main researcher-facilitator prompted participants to come up with a list of occupations they could use to test the hypothesis while researchers created a set of input images (see Figure \ref{inputs}).  Participants added occupations to a collaborative list (see Table \ref{tab:occupations}) and brainstormed fill-in-the-blank prompts that could be used to situate the occupations in different contexts. Prompts were designed in a similar way to those used in other studies of generative AI images~\cite{luccioni2024stable}, and included ``A [occupation] at work'', ``A tired [occupation],'' ``A [occupation] with friends,'' and ``A happy [occupation]''. Next, the main researcher-facilitator showed the dataset of input images compiled by the researchers (see Figure \ref{inputs}). The image dataset included four sample images built into Effect House (a Black woman, an Asian man, a White woman, and a racially ambiguous man) and a set of 10 images of celebrities selected from Wikimedia Commons (Maitreyi Ramakrishnan, Tyler James Williams, Elliot Page, Jamie Lee Curtis, Jason Momoa, Harry Styles, Lupita Nyong'o, Peppermint, Karol G, and Awkwafina) with the goal of representing a gender- and racially diverse group of popular figures. Of note, four occupations and three images were not used for testing the hypothesis due to time constraints. Finally, participants spent an hour testing different prompts and images and documenting the image outputs of Effect House's image model. To test the inputs selected in the previous step, participants collaborated as a single group, selecting 25 input occupations with four different prompts for each and 12 images to test each prompt. This required running 1200 individual tests.  To keep track of the tests and scaffold the process, the researchers created a spreadsheet on Miro, an online collaborative whiteboard tool (as visualized in Figure \ref{fig:miro}), that allowed participants to download input images, upload output images, and keep track of the tests. All the tests were conducted with consistent settings to control for possible output differences resulting from those settings (Prompt Strength: 0.50, Style: Stylized II, Transition Style: Style 1). 

\paragraph{Day 3} Participants continued running tests using the prompts from the day before.  After completing all the tests, the main researcher-facilitator invited participants to form small groups to analyze the data. She encouraged them to create a table to record their analysis by breaking the data into categories and keeping track of any changes they observed between input and output images. She emphasized that they could think about how to describe any changes they observed between inputs and outputs and how to quantify these changes.

\paragraph{Day 4} The last day of the workshop was focused on creating audit reports. Participants began by brainstorming with whom they wanted to share their findings and preferred formats for sharing their findings. They then created reports in the form of videos and slideshows. Finally, they shared the reports with the group. 

\begin{table}
\caption{Workshop Activities by Day}
\resizebox{\columnwidth}{!}{
\begin{tabular}{lll}
\toprule
\textbf{Activity}                        & \textbf{Time} & \textbf{Description}                                                                                                                                                                                                       \\ \midrule
\multicolumn{3}{c}{\textit{\textbf{Day 1}}}                                                                                                                                                                                                                                           \\ \midrule
Auditing steps activity         & 40 min       & \begin{tabular}[c]{@{}l@{}}Researcher-facilitator led a an activity where participants \\ were divided into five groups and each group had \\ to come up with a skit explaining what each step \\ involved.\end{tabular} \\ \hline
Step 1: Developing a Hypothesis & 15 min       & \begin{tabular}[c]{@{}l@{}}Participants talked amongst themselves about what kind \\ of hypothesis they could create about Effect House's \\ image model\end{tabular}                                                    \\ \hline
Step 1: Sharing Hypothesis      & 20 min       & \begin{tabular}[c]{@{}l@{}}Participants shared their hypotheses with the group; \\ researcher-facilitator proposed a hypothesis that \\ encompassed several ideas presented by the youth\end{tabular}                    \\ \midrule
\multicolumn{3}{c}{\textit{\textbf{Day 2}}}                                                                                                                                                                                                                                           \\ \midrule
Brainstorming ideas for testing          & 30 min       & \begin{tabular}[c]{@{}l@{}}Participants talked about how they could potentially \\ test the hypothesis\end{tabular}                                                                                                      \\ \hline
Reviewing 5 steps of auditing               & 30 min       & \begin{tabular}[c]{@{}l@{}}Researcher-facilitator reviewed the 5 steps of \\ auditing by showing examples from participants' work\\ from the previous week\end{tabular}                                                 \\ \hline
Why do we audit?                         & 10 min       & \begin{tabular}[c]{@{}l@{}}Facilitators led a discussion that centered on why\\ and how we audit AI/ML systems\end{tabular}                                                                                       \\ \hline
Step 2: Generating prompts               & 45 min       & \begin{tabular}[c]{@{}l@{}}Participants brainstormed prompts that they can put in\\ the Effect House's prompt interface to test their \\ hypothesis. They created a list of occupations.\end{tabular}                     \\ \hline
Step 3: Testing                          & 60 min        & \begin{tabular}[c]{@{}l@{}}Participants tested the prompts on 12 different photos\\ of different people\end{tabular}                                                                                                     \\ \midrule
\multicolumn{3}{c}{\textbf{\textit{Day 3}}}                                                                                                                                                                                                                                                    \\ \midrule
Step 3: Testing                          & 90 min      & \begin{tabular}[c]{@{}l@{}}Participants tested the prompts on 12 different photos\\ of different people\end{tabular}                                                                                                     \\ \hline
Step 4: Brainstorming ideas for analysis    & 20 min       & \begin{tabular}[c]{@{}l@{}}Participants brainstormed about how they could \\ analyze data from the tests\end{tabular}                                                                                                    \\ \hline
Step 4: Conducting the analysis           & 60 min        & \begin{tabular}[c]{@{}l@{}}Participants analyzed the outputs and put their findings \\ in Miro Board\end{tabular}                                                                                                        \\ \midrule
\multicolumn{3}{c}{\textbf{\textit{Day 4}}}                                                                                                                                                                                                                                                    \\ \midrule
Step 5: Brainstorming report ideas       & 30 min       & \begin{tabular}[c]{@{}l@{}}Participants talked about how they could report their \\ audits and possible audiences\end{tabular}                                                                                           \\ \hline
Step 5: Audit report examples            & 20 min       & \begin{tabular}[c]{@{}l@{}}Facilitators showed examples of audit reports \\ participants created the previous week and examples \\ expert-led reports\end{tabular}                                                       \\ \hline
Step 5: Creating audit reports             & 90 min     & Participants made their own audit reports in groups                                                                                                                                                                      \\ \hline
Step 5: Sharing audit reports             & 30 min       & Each group presented their report to their peers                                                                                                                                                                  \\ \bottomrule
\end{tabular}}
\label{tab:activities}
\end{table}

\subsection{Research Approach and Positionality Statement}
\label{sec:positionality}
For this study, we worked with teenagers from traditionally underrepresented identities in computing. Doing such work in an ethical manner requires centering the needs of the community and extensive engagement with participants. Before conducting the study, we worked with a youth advisory board comprised of seven 15- to 17-year-olds to brainstorm learning activities. Similarly, science center educators were involved in the brainstorming sessions and reviewed the activities prior to the workshop. Our team is invested in continuing our long-standing relationship with the science center and always aims to engage sustainably and respectfully in order to do so; one of the authors has worked with participants from this center for 15+ years and another for over five.

We recognize that using technologies developed by TikTok can present a risk for minors. During the workshop, participants did not use their personal devices or personal accounts. Instead, we partnered with the science center to provide participants with project phones, computers, and TikTok accounts. The project TikTok accounts were private, and researchers as well as science center staff were present during all interactions with the activity devices. 

We acknowledge that our own identities and backgrounds impact and prepare us for the research we do. We hold identities representing at least four different racial/ethnic backgrounds and three gender identities, as well as academic backgrounds in the learning sciences and human-computer interaction (HCI). The majority of our team lives in the same city, where our participants live and where the science center is located. Our qualifications---including expertise running expert audits, teaching high school youth, and designing learning environments---prepared us to conduct this study effectively and responsibly.
% SOMETHING ABOUT THE TEAM INCLUDING MEMBERS WITH DIVERSE EXPERTISE, INCLUDING DOING RESEARCH ON ALGORITHM AUDITING, COMPUTING EDUCATION, AND CHILD-COMPUTER INTERACTION.

\subsection{Data Collection and Analysis}

During the workshop, we collected four primary sources of data: recordings of image and video artifacts participants created (e.g., pictures of brainstorming papers, audit reports in the form of videos), screen recordings of their work on project computers and phones, the actual files of the collaborative audit (a spreadsheet with inputs and outputs of 1200 tests; see Figure \ref{fig:miro}), and researcher field notes. We analyzed this data in two different ways for this paper: first, by creating a case study of how the group collectively conducted the audit, and second, by further analyzing the dataset created by participants in order to  triangulate their findings. We discuss these in more detail next. 

\subsubsection{Case Study}
 To investigate how participants collaboratively audited Effect House's image model, we constructed a descriptive case study.  We decided to use this type of case study due to the exploratory nature of the work and because this kind of case is particularly useful to describe in detail context-specific activities \cite{yin2012case}, in our case participating in an auditing PD workshop. The main objective of this type of case study is to describe a phenomenon without aiming to provide explanations for the phenomenon or considering rival explanations~\cite{yin2018case}.  Such descriptive narratives are common in qualitative CSCW and HCI research \cite{10.1145/3359174}. For example, descriptive case studies have been used in  HCI to illustrate the context-specific nature of PD research with youth~\cite{duarte2018participatory}.
 
 Descriptive case studies are often organized around a descriptive framework that focuses the analysis on specific activities or topics. In our analysis, we use the five steps of the auditing process as a  descriptive framework. We began analysis by reviewing the audit reports that participants created \cite{yin2018case}. From the reports, we noticed a range of diverse observations and heuristics for analysis (e.g., different approaches to annotate race and gender), which sparked our interest in how participants collectively conducted the audit. To create this case study, three researchers watched 25 hours of screen recordings from the four days participants spent auditing Effect House to create videologs that documented participants' activities every two minutes. These videologs were then organized using the descriptive framework. Following, we produced analytic memos documenting participants' engagement with each step of the auditing process. Finally, we triangulated our findings with researcher field notes and video recordings of group discussions and conversations. 

\subsubsection{Triangulating Participants' Findings}
\label{sec:our-audit-method}
A major question arising during our observation of participants' auditing was if the design of the workshop supported them to reach evidence-based, credible conclusions. The purpose of the workshop was to support teens in making inferences about the actual systems that they use in their everyday lives, and having them reach inaccurate conclusions could be problematic and misleading.  CSCW studies that center on learning or educational interventions often examine whether the interventions of the studies are conducive to learning \cite{paudel2024leveraging, shen2020designing, li2020successful}. One way of approaching this is by comparing how non-experts and experts or people of different expertise complete a task. This is a common approach in the CSCW literature \cite{tang2025learning, yang2021can, foong2017novice}. As such, we decided to triangulate the findings of our participants to see if we also reached the same conclusions. Triangulating participants' findings was crucial not only to examine if this is a good way to scaffold youth in algorithm auditing but also to conduct our research responsibly, as PD may have the risk of platforming and legitimizing false ideas \cite{schafer2023participatory}.

In HCI studies, triangulating findings is not uncommon as a way to increase validity in empirical research \cite{Antti2025}. At large, triangulation involves using multiple methods, investigators, and data sources to investigate a single phenomenon. \citet{denzin2017research} describes that triangulation may involve relying on different datasets, having different researchers with complementary expertise analyze the data, analyzing a phenomenon from different theoretical perspectices or using different methods to analyze the same phenomenon. In this study, we approach triangulation  by having youth and researchers analyze the same data.

We drew upon our own expertise in conducting audits \cite{metaxa2021image, mahomed2024auditing, metaxa2021auditing} and qualitatively coded the dataset created during the PD workshop along the same axes as the youth: gender, age and race. For instance, for gender, we annotated output images for evidence of gender exaggeration (outputs being more masculine-presenting, more feminine-presenting, or the same when compared to inputs), facial hair (presence/absence), and mascara or blush in output images (presence/absence). For age, we annotated for the presence of wrinkles (presence/absence) and gray hair (presence/absence). And for race, we annotated for changes in skin complexion (outputs having lighter, darker, or the same skin complexion as inputs) and hairstyle (curlier, straighter, or the same as inputs, as well as the presence or absence of fade hairstyles) in output images. Next, three authors collectively coded an initial 20 images, discussing how and why each of the codes was applied. Then, each researcher coded the same 240 images (20\% of the data), achieving 76.25\% to 92.08\% agreement across all categories but skin complexion (50.42\% agreement) and gender exaggeration (50.83\% agreement). Due to the difficulty in agreeing on codes for these two categories, we replaced them, instead coding for change in gender representation (when the person in the output image appeared to be a different gender than the person in the input) and change in racial representation (when the input and output images were perceived as belonging to different racial groups). Following, we coded the same 240 images, achieving 82.50\% agreement for changes in racial representation and 81.25\% agreement for changes in gender representation. This resulted in substantial agreement among coders with an average inter-rater reliability of Fleiss's~$\kappa =0.65$, 95\% CI (0.58-0.72) across code categories. Finally, the same three researchers coded the remaining 960 output images. The final coding scheme is provided in Appendix A.

%% file: sections/4-findings.tex
\section{Findings}
We separate our findings into two main sections, one for each research question. In the first, we describe the results of our case study, documenting participants' engagement with each of the five activities involved in the participatory design workshop. In the second, we triangulate participants' findings by presenting our own analysis of their dataset.

\subsection{Teenagers Auditing Effect House's Image Model}
In this section, we describe the five activities participants participated in, each one focused on a separate step of the auditing process. We recount how, with the support of researcher-facilitators, participants (1) came up with a hypothesis and (2) a set of inputs to test the hypothesis, (3) ran tests, (4) analyzed data, and (5) created reports to share their findings. Here we address our first research question: \textbf{How did participants engage with this algorithm auditing activity? In particular, what choices did they make, how did these choices vary across participants, and what reflections did they have throughout the process?}

\subsubsection{Hypothesis Formation}
Participants explored the tool's functioning widely; notably, some groups narrowed in on issues commonly studied in audits, like the representation of race and gender, without being guided to do so. 

Unrelated to social bias, teams explored AI behaviors like the generative AI tool's behavior on datasets of animals and its response to images in different orientations. Selena and Twyla investigated how the system processed input images of pets. Twyla noted that pets may be misrepresented in the output images, explaining that ``when I put in a dog, it made it a cat.'' Horacio experimented with different prompts that would change the color of people's clothes, noting that sometimes these did not work as expected. Ziyi noticed that the system only generated images for pictures that were ``upside up,'' and when input images included people upside down, it did not work. Such experiments are reflective of the way participants began to creatively and broadly detect and interrogate unexpected system behaviors.

\begin{figure}[ht]
 \centering
 \includegraphics[width=\linewidth]{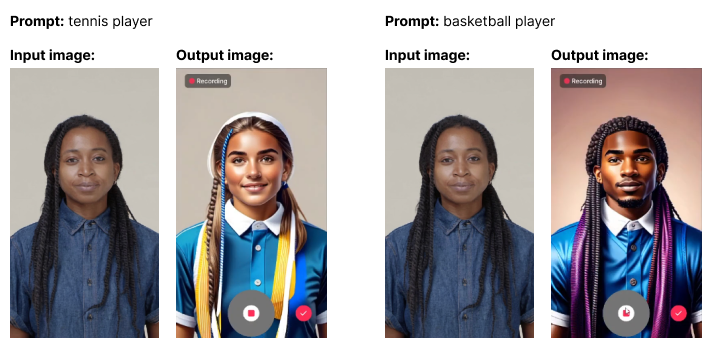}
 \caption{Inputs and outputs for Kaden's experiments with the prompts ``tennis player'' and ``basketball player.'' These experiments led the group to investigate if the image model reinforces gender and race stereotypes about different occupations.}
 \Description{Two side-by-side comparisons of input and output images. Left: The prompt is "tennis player." Input image shows a Black woman with dreads; the output image shows a tan White woman. Right: The prompt is "basketball player." Input image is the same: a Black woman with dreads; the output image shows a Black man.}
 \label{hyp}
\end{figure}

Without being prompted to do so, other groups spent their time uncovering potential issues related to the representation of race and gender, much like formal audits of such tools have~\cite{jain2022imperfect, nicoletti23}. Interestingly, some participants set out with the goal of finding such biases, while others noticed and pursued these investigations in the course of their unrelated exploration of the tool. Ibrahim and Dalia were interested in how different input images would affect the eye color of people in the images. Without explicitly focusing on race, they started to notice patterns. For instance, Ibrahim noted, for light-skinned people, ``eye color often turned green,'' and Dalia explained that she observed that if the input was a picture of a Black person, it made the eyes brown or black. One of the other groups, Ishmael and Kayden, started by playing with a scuba diving filter, explicitly looking at how it represented race and finding that it ``whitewashed'' people by giving them ``tanned skin and blonde hair.'' Then Kayden decided to modify the prompt to ``basketball player,'' trying the same prompt on a range of different images of faces. He observed that, regardless of input images, ``the skin turned Black and [the filter] gave them a beard''. He next tried the prompt ``tennis player,'' realizing that ``When I put tennis player, it made her White, but [when I did basketball player], it made her Black''. The images prompting this reflection are reproduced in Figure~\ref{hyp}. These examples show how participants identified and investigated potential harmful biases, noticing patterns and drawing upon their prior experiences~\cite{devos2022toward}. After a group discussion, participants agreed to build on Ishmael and Kayden's work to investigate the following hypothesis: \textbf{Effect House's image model reinforces gender and race stereotypes about different occupations}. 

\subsubsection{Designing Inputs}

While deciding on the occupations they could use to test their hypothesis (see Figure \ref{tab:occupations}, participants took different approaches. As we will describe, some did so through discussion, reflecting on their own experiences and perceptions of stereotypes, while others were more hands-on, using Effect House to conduct preliminary explorations and tests of their ideas. 

Some groups of participants reflected on common stereotypes related to occupations and how these shaped their expectations of the outputs that Effect House would generate. Kalem suggested ``chef'' as an occupation because ``usually chefs are White and European.'' Ishmael interjected, saying that thinking of all chefs as White was ``just racism.'' Kalem responded by explaining that he thought they were meant to find stereotypes and test if the systems created images based on those stereotypes. Kalem wrote down ``teacher,'' explaining he would expect outputs to look like White women. Ishmael proposed ``rapper'' because ``it makes me think of Black men, specifically Tupac.'' Participants also reflected on their own personal experiences with race and occupations. Kalem shared that he thought ``nurse'' would be a good occupation to investigate because his mother is a nurse and all of her friends are also nurses, and they are all Black women---anticipating a similar trend might be reflected when the filter was used on a range of different faces. Notably, this particular group brainstormed through discussion and did not try to empirically explore any of these occupations on the platform. Like participants in everyday audit studies~\cite{devos2022toward}, participants drew on their prior personal experiences and knowledge of societal biases to generate inputs.

At a different table, a group used Effect House to test different occupations as they discussed them. Twyla, tested ``basketball player'' on an input image of a White woman, observing that the result resembled a Black woman. Horacio explained that the group had to consider traditionally masculine jobs like ``soldier'' to see the outputs Effect House would generate. Twyla agreed, saying that she always thinks of war as something associated with men. She suggested ``construction worker,'' ``chef,'' and ``cook'' as occupations to try. She explained that while chefs are usually associated with men, cooks are often with women, ``they just do the same job.'' Similarly, Ibrahim used Effect House to come up with possible occupations. He tested ``basketball player,'' ``computer scientist,'' ``gardener,'' ``7/11 worker,'' ``chef,'' and ``teacher'' on four different images.

% occupations and prompts here
\begin{table}
\caption{List of occupations created by participants to test their hypothesis.}
\begin{tabular}{llll}
\toprule
\multicolumn{4}{l}{\textbf{Occupations}}                         \\ \hline
1. Tattoo artist    & 9. New anchor & 17. Taxi driver   & 25. Receptionist      \\
2. President of the USA & 10. Scammer  & 18. 7/11 worker   & 26. Pizza delivery person* \\
3. Carpenter      & 11. Rapper  & 19. Lawyer     & 27. Tech support*     \\
4. Construction worker & 12. Judge   & 20. STEM student  & 28. Oil salesman*     \\
5. Priest        & 13. Senator  & 21. Nail technician & 29. Corner store worker*  \\
6. Fast food worker   & 14. Janitor  & 22. Harvard Student &              \\
7. Basketball player  & 15. Astronaut & 23. Mathematician  &              \\
8. Teacher       & 16. Chef   & 24. Music artist  &              \\ \bottomrule
\multicolumn{4}{l}{\textit{*Occupations 26-29 were not used in testing and analysis}} 
\label{tab:occupations}
\end{tabular}
\end{table}

\begin{figure}[ht]
 \centering
 \includegraphics[width=\linewidth]{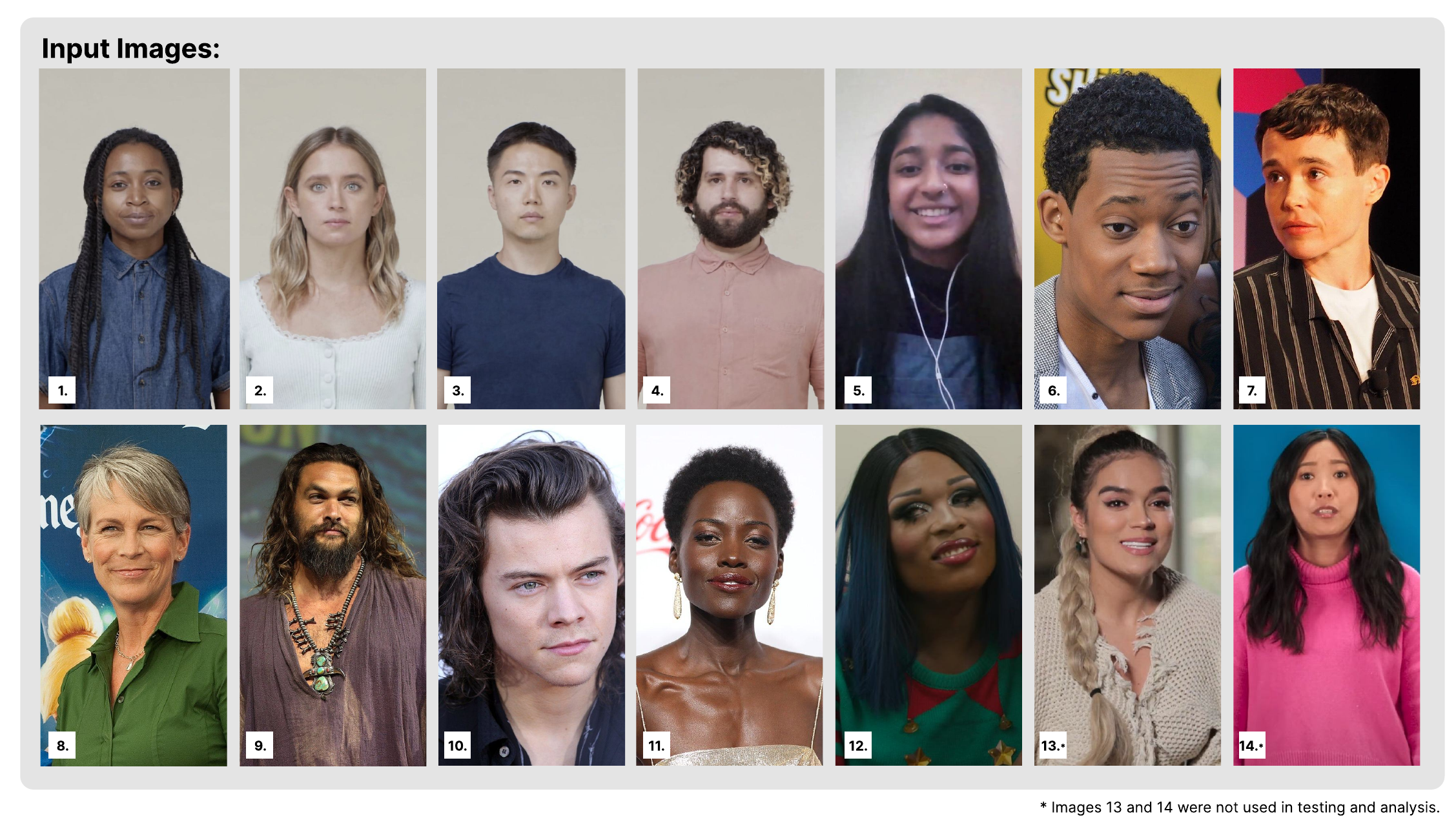}
 \caption{The input images used by participants in their audit, including 4 default Effect House inputs and 10 images of celebrities selected from Wikimedia Commons.}
 \Description{14 headshots used as input images. Top Row: A Black woman with long dreads, a White woman with shoulder-length blonde hair, an East Asian man with short hair, a racially ambiguous man with curly hair and highlights, Maitreyi Ramakrishnan, Tyler James Williams, Elliot Page. Bottom Row: Jamie Lee Curtis, Jason Mamoa, Harry Styles, Lupita Nyong'o, Peppermint, Karol G, Awkwafina}
 \label{inputs}
\end{figure}

\subsubsection{Running Tests}

\begin{figure}[ht]
 \centering
 \includegraphics[width=\linewidth]{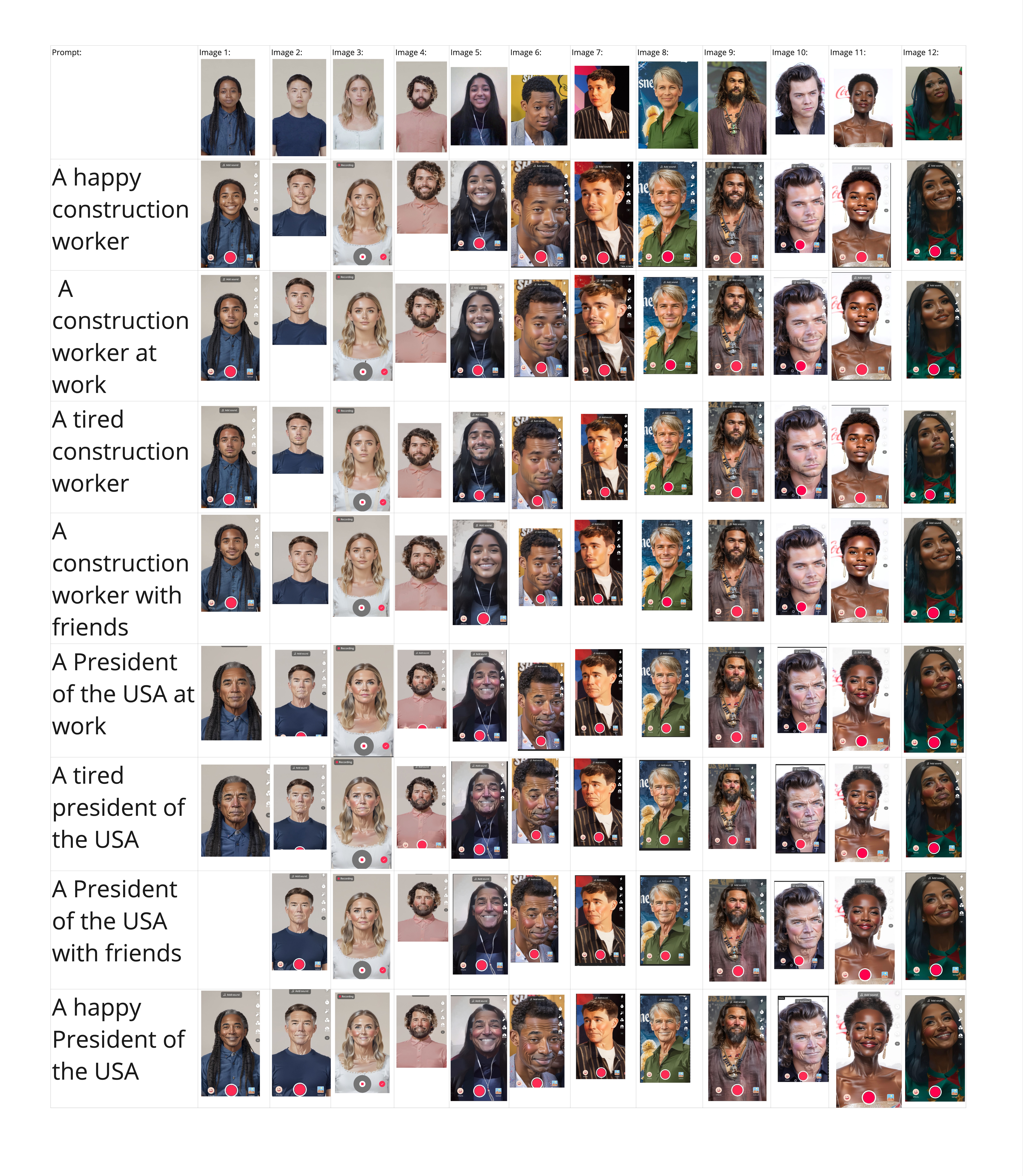}
 \caption{Screenshot of a section of the spreadsheet where participants kept track of their tests.}
 \Description{Screenshot of a spreadsheet with 12 input images along the horizontal axis and 8 text-based input prompts along the vertical axis. The table displays the output images from Effect House and TikTok.}
 \label{fig:miro}
\end{figure}

Running the tests provided participants with opportunities to make inferences about the outputs, notice patterns, and reflect on their own perceptions of occupations. Ziyi, aware of the repetitive nature of running tests on all the images (see Figure \ref{inputs}), explained to a science center instructor that the trick was focusing on ``one at a time.'' She was also continuously reflecting and noticing patterns while running tests. For example, when testing the construction worker prompts with different images, she noted, ``they all look the same.'' Later, while running tests for the nail technician, she explained that she was unsurprised that they all looked very feminine because her mom is a nail technician, and she noticed that it was a very feminine occupation. Similarly, Dalia reflected after running tests for the occupation ``tattoo artist'', noting ``a lot of people got tattoos... I wonder why [Effect House] gave her, like, a neck tattoo''.

Even while running the tests and adding output images to the shared spreadsheet (see Figure \ref{fig:miro}), participants continued to playfully experiment with Effect House. One participant, Kayden, took the opportunity to further play with the parameters and test images using his own name as a prompt. For instance, after testing a picture of Jason Mamoa with the prompt ``a president of the USA at work,'' he decided to modify the prompt using his own name (e.g., ``Kayden Lastname at work'') to see what came up. The output was an image of a young Black man. He then tried his own name as a prompt with an image of a Black woman and got an output where the woman had more masculine features, and her race remained unchanged. While these tests were not documented on the shared spreadsheet and did not contribute to the overall audit, screen recordings revealed how the testing activity inspired him to go beyond the task he was supposed to be completing. 

\subsubsection{Analyzing Data}
Participants took different approaches to analyzing the data, with some describing the changes they observed, others annotating perceived gender and race of images, and some centering on concrete observable attributes. Participants also annotated age-related features, even though this attribute was not planned in the original hypothesis (about gender and racial stereotypes). At the same time, while analyzing data, participants continued to further experiment with the system by running more tests. 

Some teens, including Twylia, decided to describe the visible differences between each input and output image. For example, for the prompt ``Janitor at work'' and image 1 (an image of a Black woman with dreads), she noted that the output had ``visible makeup, bigger lips, toned eyebrows, and a smoother face.'' A researcher-facilitator approached Twylia and suggested she count some of the changes she was noticing across images rather than just qualitatively describing them. Based on her interpretation of the suggestion, Twylia continued with descriptive annotations of the output images and also began assigning a masculinity and femininity score, ranging from zero to a hundred, to each output image.

Other participants labeled the data with descriptors based on their own overall perception of the output images. Ibrahim, Kalem, Dalia, and Taylor, for example, annotated whether the people in output images looked more feminine or masculine, older or younger, and darker- or lighter-skinned than the inputs. Similarly, Ishmael and Brooklyn Mae annotated the input and output images using defined binary gender and age categories such as male/female and young/old.

One group of participants decided to focus on observable features rather than their own perception of the outputs. Motivating this choice, Selena explained that ``what `older' means is kind of subjective.'' Ziyi also emphasized the need to specify what ``older'' meant because it was hard to know and estimate the age of the AI-generated images. She suggested that they should look at the output images to see if these had wrinkles and gray hair in order to concretely measure how many of the output images looked older. In a similar episode, Ziyi and Kalem discussed how to account for gender changes in the representation of rappers. First, they considered counting how many of the inputs were men and comparing them with the number of output images that (in their own perception) were men. But Ziyi again suggested finding a more concrete way to label the images, focusing on a specific physical attribute like the presence or absence of facial hair.

While analyzing data, one participant, Horacio, decided that he needed to do further testing. He started his analysis by annotating what he noticed in each output image with notes like ``gender stays the same but face is different.'' Unprompted, he decided to rerun some of the tests and change the prompt strength from 0.5 to 1.0. He noted that for some prompts, he did not observe a change in gender between input and output images at a strength of 0.5, such changes were noticeable at 1.0 strength. The process of analyzing led Horacio to reconsider how other variables in the auditing process could affect the outputs and to loop back to the data collection step to experiment with one variable.

\begin{figure}[ht]
 \centering
 \includegraphics[width=\linewidth]{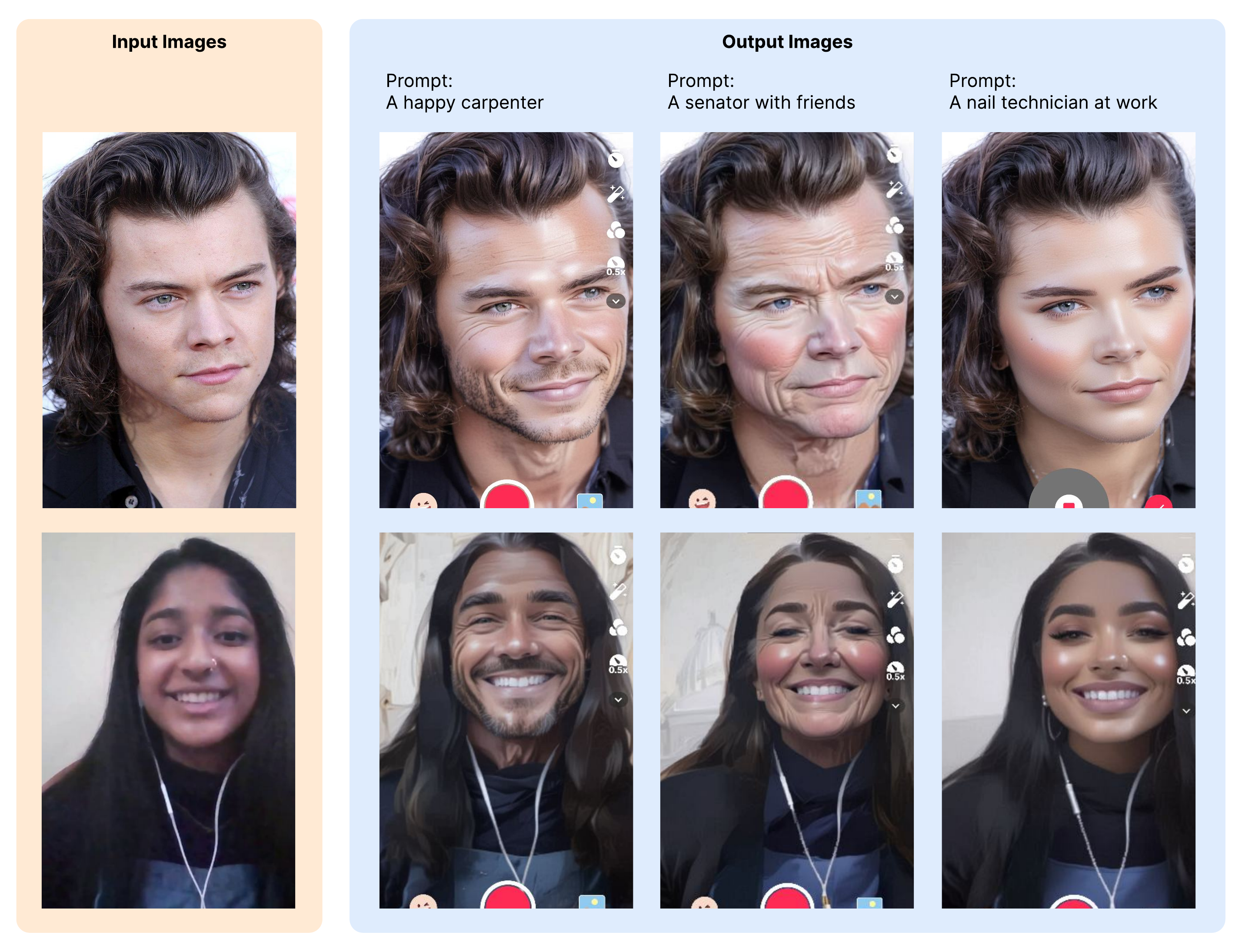}
 \caption{Examples of input and output images for three prompts.}
 \Description{Images of Maitreyi Ramakrishnan and Harry Styles transformed using the prompts "a happy carpenter," "a senator with friends," and "a nail technician at work."}
 \label{fig:inout}
\end{figure}

Overall, participants analyzed 13 out of the 25 occupations that were collected in the previous step (see Table \ref{findings} for findings by group). They annotated the image pairs for changes in representation of gender, age, and race caused by the model by comparing the input and output images. Here we provide examples of participants' findings for specific occupations. Four groups analyzed the gender representation in output images for nail technician prompts (for an example, see Figure \ref{fig:inout}), noting that the outputs were more feminine, with some participants arguing that all images looked more feminine (Kayden and Dalia; Taylor), and one that 87\% looked more feminine (Ibrahim). Participants that looked at specific feminine-coded features noted the presence of makeup in 83\% of the nail technician outputs (Kalem, Selena and Ziyi). Two groups analyzed the outputs for judges, noticing that judges in general had more feminine characteristics (Ishmael and Brooklyn Mae; Taylor). In terms of age, two groups also looked at judges, noting that 24\% of the faces in the output images were young (Ishmael and Brooklyn Mae) and 75\% had wrinkles (Kalem, Selena, and Ziyi). They made similar observations for other occupations, noting that fast food workers looked young, while outputs for senators and janitors looked older. Less attention was given to race, with two groups noting race-related attributes: that all rapper outputs had darker skin than the input images (Ibrahim), and that senators had ``White aesthetics like blue eyes and blond hair'' (Twyla and Horacio).

\begin{table}
\caption{Findings from small group analyses of the data generated in the auditing process.}
\resizebox{\columnwidth}{!}{
\begin{tabular}{ll}
\toprule
\textbf{Group}          & \textbf{Audit Findings}                                                                                                                                                                                    \\ \midrule
\textit{Kalem, Selena and Ziyi}  & \begin{tabular}[c]{@{}l@{}}Nail techs: 83\% have makeup {[}proxy for feminine features{]}\\ Fast food worker: 100\% of outputs had smooth skin {[}proxy for younger age{]}\\ Rapper: 95\% of outputs had goatees or facial hair {[}proxy for masculine features{]}\\ Judge: 75\% had wrinkles {[}proxy for older age{]}\end{tabular}                                     \\ \hline
\textit{Ibrahim}         & \begin{tabular}[c]{@{}l@{}}Rapper: 100\% of outputs had darkened skin; 96\% looked more masculine\\ Nail tech: 87\% looked more feminine \\ Taxi driver: 56\% looked older\end{tabular}                                                                                                 \\ \hline
\textit{Ishmael and Brooklyn Mae} & \begin{tabular}[c]{@{}l@{}}Priest: 100\% outputs were men, even though inputs were 50/50 split\\ Judge: 100\% of outputs appeared female/femme\\ Judge: 24\% outputs were young\end{tabular}                                                                                              \\ \hline
\textit{Twyla and Horacio}    & \begin{tabular}[c]{@{}l@{}}Senator: 100\% were old; most had white aesthetics like blue eyes, blonde hair, \\ and older (with wrinkles added); 95\% were men\\ Fast food worker at work: tone eyebrows, smoother face, visible makeup; \\ 80\% looked feminine; 20\% looked masculine\\ Janitor at work: 85\% looking older; 80\% looking masculine; 20\% looking feminine\end{tabular} \\ \hline
\textit{Kayden and Dalia}     & \begin{tabular}[c]{@{}l@{}}Carpenter: 91.7\% of outputs were men\\ President: made everyone older, added wrinkles and gray hair\\ Tattoo artist: added tattoos on bodies like faces\\ Stem scholars: made everyone look younger\\ Nail technician: made everyone look more feminine\end{tabular}                                            \\ \hline
Taylor              & \begin{tabular}[c]{@{}l@{}}Nail techs: outputs images were more feminine; sometimes added makeup\\ Receptionists: more feminine characteristics \\ Judges: more feminine characteristics\end{tabular}                                                                                          \\ \bottomrule
\end{tabular}}
\label{findings}
\end{table}

\subsubsection{Creating audit reports}

Participants decided on different audiences for their reports and different ways to communicate their findings. They considered distinct relevant parties\footnote{We use the term ``relevant parties'' rather than ``stakeholders''; for a discussion on the reasoning behind this decision, see~\cite{reed2024reimagining}.} including the engineers and others involved in the development of Effect House, users that create filters on Effect House, and users of filters on TikTok. Kalem suggested that it was important to share the findings with the developers of Effect House so they might improve the generative model. Ibrahim and Twyla argued that the findings should be shared with filter designers so that they could consider the biases that the system introduces even when the filter designers might not intend them. Ibrahim explained, ``Effect House can be biased no matter what''. Dalia suggested sharing the findings with other teenagers who use TikTok because ``AI is everywhere and young people don't always know how it works.'' She explained that without knowing how systematic some of these issues are, other teens could find themselves thinking, ``Why is it doing this? Why is it doing this to me?'' 

Some groups decided to make their own TikTok videos to report their audit findings; one group, Twyla and Horacio, decided to make a slide presentation. In a different group, Ziyi and Kalem wrote a script for a TikTok video directed towards the developers of Effect House. Ziyi emphasized that it was important to explain how and why they selected the different occupations they analyzed, noting that they chose to focus on occupations that they perceived as reflecting ``the most physical change'' between the input and output images. Kalem explained that he hoped the developers of Effect House could take this feedback and make the platform more inclusive so that users could have access to better filters on TikTok.

\subsection{Triangulating Participants' Findings}

%Expert audits generally involve rigor and exhaustiveness in the selection of inputs and also in the analysis of outputs, and doing so at scale is a challenge that auditors must confront. Solutions vary widely; for example, prior work investigating gender and racial representation of occupations in AI-curated or -generated images has been analyzed by using crowdsourced human annotation~\cite{metaxa2021image}, automated annotation using computer vision~\cite{naik2023social}, or having transformer models create captions for the images~\cite{luccioni2024stable}. 
% Prior research on gender and racial representation of occupations in image search also used human annotation of images that could then be compared to a baseline (often nation-wide labor statistics)~\cite{metaxa2021image}. 
To address our second research question, \textbf{Did the workshop support participants to reach evidence-based, credible conclusions in their audit?}, we triangulated participants' analyses. We built on our team's prior experience in algorithm auditing by conducting our own analysis of the dataset created by youth. (For details about how we conducted this analysis, see Methods - Section \ref{sec:our-audit-method}). The goal of this analysis was to answer the hypotheses developed during the workshop to investigate if outputs from Effect House's image model reinforced gender and race stereotypes about different occupations and use these results to triangulate participants' audit results. In addition to gender and race, we also added age representation since several groups of participants also analyzed this kind of bias. 

\begin{figure}[h]
 \centering
 \includegraphics[width=\linewidth]{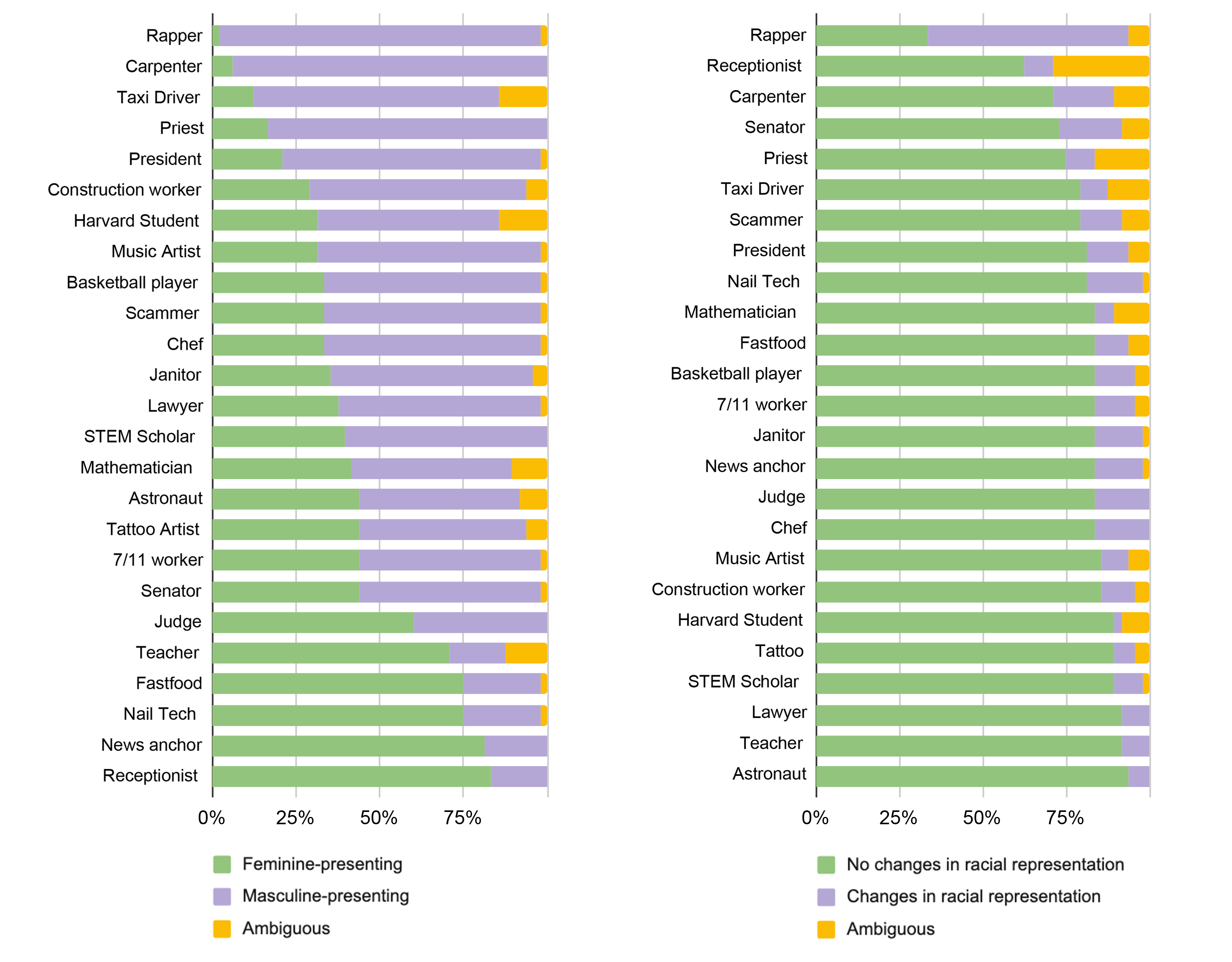}
 \caption{Bar graphs depicting the gender representations (left graph) and the changes in racial representations (right graph) of the output images for each occupation.}
 \Description[Two bar graphs depicting the gender representations (left graph) and the changes in racial representations (right graph) of the output images for each occupation.]{2 bar graphs. The left chart shows the percentage of feminine-presenting, masculine-presenting, and ambiguous outputs. The occupations: receptionist, news anchor, nail tech, fast food worker, teacher, and judge had higher feminine-presenting outputs. The right chart shows the percentage of changes in racial representation from the input image. The rapper occupation prompt changed the racial representation for over 50 percent of the input images.}
 \label{fig:chart}
\end{figure}

\paragraph{Gender Representation}
Our analysis of the dataset, created during the collaborative audit, revealed that Effect House generated a greater number of masculine-presenting outputs for 19 out of the 25 occupations (see Figure \ref{fig:chart}), depicting an overrepresentation of masculine-presenting figures in these occupations. Here we describe our findings in relation to participants' findings. The input images were selected to be balanced in gender representation (50\% F, 50\% M), but after our own annotation, the outputs were 41\% F, 55\% M, 4\% ambiguous (A). In certain occupations, the gender imbalance was more pronounced. For instance, 96\% of the output images for rappers appeared masculine-presenting, followed by 94\% for carpenters, 83\% for priests, 73\% for taxi drivers, and 77\% for presidents. On the other hand, 83\% of receptionists were feminine-presenting, followed by 81\% for news anchors, 75\% for fast food workers, and 75\% for nail technicians. We observed changes in gender representation in at least one test (out of 100 tests) for all input images, with the exception of the image of Jason Mamoa and image 4 (racially ambiguous man with a beard), who always retained a stereotypically masculine presentation in the output image regardless of the input prompt.

Although researchers and participants obtained slightly different quantitative results, the conclusions drawn from both analyses were similar, namely that---in terms of gender representation and biases---Effect House's filter masculinized input images when prompted with occupations like rapper, carpenter, priest, and janitor and feminized outputs for occupations like fast food workers, receptionists, and nail technicians. For example, researchers found that 94\% of output images for carpenters were masculine-presenting, while Kayden and Dalia reported that 91.7\% of output images were men. In other occupations, such as fast food workers, researchers coded 75\% of output images as feminine, while participants argued 80\% looked feminine. Participants analyzed the images of rappers in different ways. Kalem, Selena, and Ziyi used facial hair as a proxy for masculine-presenting outputs, finding that 95\% of the outputs for rappers had facial hair, while Ibrahim directly labeled the images as masculine or feminine in his own perception, finding that 96\% of the outputs were masculine-presenting. Our analysis concorded with Ibrahim that 96\% of outputs were masculine-presenting and came close to Kalem's group; we found that 100\% of outputs for rappers had facial hair (including stubble).

\paragraph{Racial Representation}
Participants' analyses of racial representation were limited, with only two groups addressing racial representation. Ibrahim reported that 100\% of outputs for rappers had darker skin than the input images, while Twyla and Horacio noted that most outputs for senators featured "White aesthetics," like blue eyes and blonde hair. In our analysis, we compared input and output images, annotating whether the race of the person in the output image appeared (subjectively) to differ from that of the input. Our analysis revealed that changes in racial representation were not as prominent as those in gender representation. Overall, when comparing input and output images, we found that racial representation changed in 13\% of cases, with changes occurring across all occupations. The occupations with the highest frequency of change in racial representation were rappers (60\%), carpenters (19\%), and senators (19\%). Occupations with the lowest frequency of change in racial representation included astronauts (6\%), lawyers (8\%), and teachers (8\%).

% Changes in racial representation were most noticeable for two input images: Effect House's default ``image 2'' (an East Asian man) and the image of Maitreyi Ramakrishnan (a South Asian woman). For image 2, we observed changes in racial representation in 89\% of cases. With most occupations, image 2 became a White man with the exception of rapper. For images of Ramakrishnan, these had changes in racial representation in 38\% of cases; outputs included being transformed into White men (e.g., output for ``a tired judge'') and Black men (e.g., output for ``a happy rapper''). It is important to note that these were the only two images of Asian people in our dataset.

Given that these results were less stark than those related to gender, it is possible that the participants did not focus as much on this form of bias because they empirically found the patterns in racial representation less pronounced and interesting in their experiences running the tests. 
% In our analysis, we confirmed Ibrahim's finding that most output images for rappers looked like Black men, with 60\% of outputs showing changes in racial representation. For senators, however, we only reported race changes in 19\% of the outputs. We noted race changes for image 2 (East Asian man), image 4 (racially ambiguous man) and Jason Mamoa; the outputs of these three images appeared to be white men. 

\paragraph{Age Representation}
While age representation was not a part of the initial hypothesis, it emerged as a key theme in participants' audits, with 5 out of 6 groups including observations about changes in age representation. We evaluated age representation using two visual indicators: wrinkles and gray hair. A change in age representation was identified if at least one of these changed between the input and output images. 

We found that changes in age representation were more prominent than changes in racial and gender representation, with changes in age representation occurring in 46.75\% of cases. Notably, changes in wrinkles were more prominent than changes in gray hair, and the addition of features in the output image was more common than their removal. The presence of wrinkles changed (added or removed) in 44.67\% of images, including 40\% that involved the addition of wrinkles. Gray hair changed in only 10.33\% of images. The occupations that showed the most change in age representation were priest (83.3\%), president (83.3\%), and senator (75\%). Senators had the highest incidence of wrinkles in output images (100\%), followed by news anchors (95.8\%), presidents (95.8\%), taxi drivers (93.7\%), and carpenters (93.7\%). Occupations with the highest incidence of wrinkle removal were STEM student (16.67\%), nail technician (14.58\%), and fast food worker (12.5\%).

Participants' analyses were similar to the researchers' findings, including that Effect House tended to age input images when prompted with certain occupations—such as president, senator, and taxi driver—and reduced the perceived age of images when prompted with STEM student. However, participants employed distinct strategies to analyze images, with some (Kalem, Selena, and Ziyi) focusing on specific visual markers, such as the presence or lack of wrinkles, as proxies for age, and others (Kayden and Dalia) taking a comparative approach, describing output images as ``older'' or ``younger'' than the inputs. A third approach by Ishmael and Brooklyn Mae relied on personal interpretations of ``young'' and old'', as they described in their audit that 24\% of outputs for judges were ``young''.

%% file: sections/5-discussion.tex
\section{Discussion}
 
At the highest level, this study contributes evidence that, beyond recognizing isolated instances of representational bias and harm, teenagers can effectively engage in algorithm auditing, collaboratively and \textit{empirically} investigating potentially harmful behaviors in real-world algorithmic systems. Our study confirms evidence from prior research that young people build on their everyday experiences to identify potential biases \cite{morales2024youth, solyst2023potential} and extends this by showing that when participating in scaffolded PD activities, teenagers can lead full-fledged audits to empirically research systemic bias and harm. In this section, first we discuss participants' approaches to auditing in relation to auditing literature, then we focus on the implications of our findings for future auditing activities involving youth. Subsequently, we also reflect on some implications for the field of auditing, especially as everyday perspectives are beginning to be included in various ways. Finally, we describe key limitations to be addressed and other future directions in which we hope our team and others will expand this line of research.

\subsection{Teens Conducting an Algorithm Audit}

While prior research shows that young people can identify harmful biases in researcher-selected scenarios ~\cite{solyst2023potential, morales2024youth}; our case study demonstrates that youth can be protagonists ~\cite{iversen2017child} or the main decision-makers in auditing a generative AI system from beginning to end. The PD workshop was structurally open ~\cite{zacklad2003communities, bannon2012design}, providing teens with autonomy to define what they wanted to investigate. The five steps served as a basic structure that supported participants to be the main agents driving the auditing process ~\cite{iversen2017child}. In the following paragraphs, we discuss how teens navigated each of the steps.

Participants developed hypotheses based on experimentation and personal experiences, not only noticing unexpected behaviors but also potentially harmful biases. While some expert audits are conducted as follow-ups to well-documented problematic system behaviors~\cite{metaxa2021image} and others are initiated in compliance with local laws~\cite{groves2024auditing}, expert audits have also been motivated by the researchers' own personal experiences interacting with sociotechnical systems~\cite{buolamwini2018gender, sweeney2013discrimination}. The open-ended exploration of the tool and its behaviors was instrumental in helping participants develop hypotheses that were relevant to their interests, identities, and experiences.

In most expert audits, auditors create a set of inputs that are systematic, thorough, and thoughtful, and that can be used to rigorously test the hypothesis. It is notable that the domain of common occupations, selected by the participants without explicit direction from researchers, is a popular framing for expert audits. For example, previous studies on gender and race representation of occupations in image searches and the outputs of generative models have used the US Bureau of Labor and Statistics (BLS) categorization of occupations as a starting point~\cite{kay2015unequal, metaxa2021image, bianchi2023easily, naik2023social, luccioni2024stable}. Participants were less exhaustive but very creative when they selected their list of occupations. Like Kalem, when thinking about the gendered and racialized stereotypes around nursing in relation to his mother, participants built on their own personal experiences. They also reflected on societal biases they observed in their everyday lives and suggested occupations such as rapper and nail technician that, to our knowledge, have not been included in previous expert-led studies.

Repeatedly querying an algorithmic system using thoughtfully designed inputs and recording outputs can be tedious and time-consuming. Despite its tedium, we observed that participants were generally engaged in running the tests, often performing some informal analysis of outputs while testing and reflecting on their perceptions of stereotypes. Sometimes participants deviated from the task at hand to pursue their own open-ended investigations. These examples show that auditing with teenagers can be less structured than expert auditing , with learners engaging with different steps of the auditing process at the same time.

Participants adopted various methods to analyze their data, recording observed changes, annotating perceived gender and race of photos, and focusing on observable features. We were able to triangulate their findings to confirm that Effect House generated a greater number of masculine-presenting outputs for 19 out of the 25 occupations, and that age representation reflected societal stereotypes.

Like expert auditors who report their findings with the goal of effecting change ~\cite{mahomed2024auditing, nicoletti23}, participants in our study had a clear understanding that different relevant parties needed to access the findings for different purposes and created distinct messages directed to these parties.

The PD workshop successfully scaffolded participants in conducting an audit. Our study further highlights the promise of expanding the roles young people can play in PD to involve them as auditors of systems that are relevant to their daily lives ~\cite{iversen2017child, morales2024youth}. For child-computer interaction research, positioning teenagers as auditors of the technologies that are designed and marketed towards them is particularly important, as they may be able to identify issues that designers, adults, or experts cannot find.

\subsection{Implications for Future Auditing Learning Activities}
Our first research question asked about how participants engaged in this auditing activity. Reflecting on the findings of this first research question, we consider the adaptation and scaffolding of the auditing process promising for future activities with teenagers, and the algorithmic justice topic suitable and resonant with this group of teens. The activity clearly resonated with participants; they connected with their own lived experience throughout the process. For example, Kayden tried running several prompts with his own name. We saw this clear personal engagement even when directly addressing complex topics like social biases, as when Twyla discussed the differences between stereotypes of chefs and cooks. Participants also demonstrated their creativity. Different groups conducted parallel steps of the audit very differently---for example, generating distinct ways to evaluate age representation, such as by subjectively comparing inputs and outputs or by looking for the presence of wrinkles and gray hair. 

Our second research question asked whether the workshop supported participants to reach evidence-based, credible conclusions in their audit. This is especially concerning because this activity was meant to allow them to draw conclusions about real systems they use in their daily lives. It could be problematic to leave them with inaccurate conclusions, something we could not control in advance as the process unfolded organically. By triangulating participants' findings, undergirded by prior experience in expert auditing, we observed that their findings were quite close to our own---for example, finding that Effect House appears to produce outputs that are systematically biased according to social stereotypes of different occupations. Although they did not analyze the full dataset, as our team did after the workshop, and although the precise numbers they calculated varied from ours in some cases (e.g., while we found that 75\% of output images for fast food workers were feminine-presenting, Twyla and Horacio argued 80\% looked feminine), the overall direction of the trends they reported matched ours. This is evidence that the PD workshop supported participants in reaching evidence-based, credible conclusions. 

%In one notable case---deciding whether to annotate biases by marking whether the output image was ``more masculine'' or simply had some specific attribute change (e.g., gained a beard)---participants' own struggles prognosticated our own. While some groups of participants, such as Kayden and Dalia, opted to annotate based on their own overall perceptions of the image, others, such as Kalem and Ziyi identified key characteristics like facial hair and wrinkles to annotate instead. This decision to shift to more easily measured attributes was a very thoughtful innovation reminiscent of the kinds of choices researchers and scientists often have to make in order to concretely measure their phenomena of interest. 

Moreover, several of the participants' divergences from formal auditing processes proved insightful and valuable. When analyzing the data, for example, several groups began to notice trends related to age, diverging from the hypothesis. Age biases are not a common topic for expert audits; as reflected in the related work we described in Section \ref{sec:relwork}, we do not know of any other image bias audits that study age. But clearly this dimension was salient to participants, who uncovered notable trends that complemented the findings on race and gender (e.g., that filter outputs for presidents, in addition to looking more masculine-presenting, also looked older). Similarly, when selecting the occupations to examine, they chose several uncommon ones not previously explored in the expert literature, like nail technician and rapper, that nevertheless produced interesting findings. 

Looking ahead to future deployments of auditing-based learning activities, we encourage educators to integrate opportunities for experimentation throughout the process and not restrict learners very strictly to the task at hand. We observed that the participants' approaches were playful, as they devised alternative hypotheses and conducted impromptu open-ended explorations throughout the process. We saw them gaining inspiration in later activities that prompted them to loop back to earlier stages of the audit and try new directions. And as we just described, although efforts should be made to ensure that learning activities will leave teenagers with valid and well-informed conclusions, there are many other dimensions in which learners follow their creative impulses and make different decisions than what expert auditors would normally do. 

\subsection{Implications for the Field of Auditing}

In addition to contributing to the literature on child-computer interaction, we see a couple of implications of this work for auditing research. First, it confirms the value of participatory approaches to auditing that involve non-expert real system users in shaping the questions, methods, and interpretations of an audit~\cite{lam2022end}. Notably, most prior work in this space has focused on the potential for users to pose questions and conduct explorations. Beyond these steps, we also saw participants consider how their findings should be communicated to different relevant parties in distinct ways (e.g., making TikTok videos to communicate findings to TikTok users). This suggests that user populations may be especially well-positioned to effectively communicate their findings to peer groups after conducting an audit (the fifth step in our set of activities). We hope subsequent work in end-user auditing will consider ways to leverage end users' expertise and trust among their peers to not only conduct audits but also communicate audit results. 

And of course, this work expands the idea of end-user auditing to include teenagers. Young people, as avid users of many AI systems, have important expertise that we saw arise in this study (e.g., the consideration of age biases, the choice of occupations that resonated with them personally). Like other audits involving everyday users have shown~\cite{shen2021everyday}, our 14-15-year-old participants contributed valuable auditing insights based on their lived experiences. 

%%here bring the DeVrio FAccT paper 

\subsection{Limitations and Future Work}
Perhaps the largest limitation of this research is that we only present an analysis of one case study of a single algorithmic system with a specific group of teenagers. Given the value we saw for teenagers, we believe it is worth expanding on this work to develop materials for conducting workshops focusing on other platforms, other age groups, and additional groups of learners. We encourage other researchers to do so and also plan to do this ourselves, and, as we referenced in our Positionality Statement (Section \ref{sec:positionality}) we emphasize the importance of partnering with teenagers and collaboratively creating appropriate learning materials with them. 

We conducted these activities in the context of an extended summer program, but the concept of engaging teenagers in auditing could be more impactful if designed for classroom settings with a wider range of learners---not just those with interest and access to an extracurricular program, like the participants in our study. To this end, our next steps involve exploring how the five auditing steps that framed our activities could be used to integrate algorithm auditing activities in classroom settings. This will raise some interesting challenges, for instance, pertaining to many schools' blocking of platforms like TikTok on school networks and use of mobile devices in classrooms. To address these issues, we must take a participatory approach and partner with educators and teenagers to explore other possibilities.

Focusing specifically on the design of the PD workshop, we want to flag a couple of challenges. Creating a large data set to conduct a full-fledged audit was time-consuming and tedious. In adapting these kinds of activities to classroom settings, we could consider focusing on analysis and providing teenagers with existing curated datasets. Another challenge was the impossibility of the participants to annotate all 1200 tests, an intractable number for any learning activity of reasonable length. To address this, it is important to design tools that can streamline and support annotation and to consider other pedagogical designs, such as having participants annotate some of the data while providing them with already annotated data. Finally, after reflecting on the workshop ourselves, we also wondered whether adding a sixth step, Reflection, in which participants could reflect on their experiences and challenges, might support learners in thinking about how auditing practices can be applicable in their everyday lives.

% In this context, we also want to acknowledge the central role that the workshop facilitators played in engaging the youth teams in conducting their algorithm audits, and by extension teachers who want to integrate auditing into their classroom activities.

%% file: sections/6-conclusion.tex
\section{Conclusion}
Motivated by the importance of fostering artificial intelligence and machine learning (AI/ML) literacies among youth, we drew on algorithm auditing research to design and deploy an auditing-based participatory design workshop with a group of 14 teenagers. Auditing is a method used by experts to interrogate algorithmic systems and their social impacts that has more recently seen extensions through which everyday users (generally adults) can participate in similar processes in the context of their daily interactions with these technologies. The teenagers in our study (ages 14–15) engaged in a PD workshop auditing the generative AI model that powers TikTok filters using Effect House, TikTok's filter development environment. Our case study shows that teenagers are able to conduct a complete and collaborative audit of a real-world algorithmic system when scaffolded appropriately. Furthermore, conducting our own analysis, we were able to triangulate their findings, confirming the potential of the workshop in scaffolding youth to reach evidence-based, credible conclusions. This work demonstrates the potential for researchers to partner with teenagers in future audits of algorithmic systems that young people use in their daily lives and also to design AI literacies learning activities that support young people in conducting algorithm audits of the systems that they interact with every day.